\documentclass[fleqn,a4paper,11pt]{article}

\usepackage{amsmath, amssymb, amsthm, bbm, lscape, mathrsfs, sectsty}
\usepackage{graphicx,color}

\usepackage[hang, small, bf, margin=20pt, tableposition=top]{caption}
\newtheorem{thm}{Theorem}

\newtheorem{lem}[thm]{Lemma}

\numberwithin{equation}{section}

\newcommand{\bm}[1]{\mbox{\boldmath{$#1$}}}

\newcommand{\bell}{\bm\ell}

\newcommand{\be}{\bm e}
\newcommand{\bg}{\bm g}

\begin{document}

\title{A classification of the symmetries of uniform discrete defective crystals}

\author{Rachel Nicks }
\date{}
\maketitle

\begin{abstract}
Crystals which have a uniform distribution of defects are endowed with a Lie group description which allows one to construct an associated discrete structure. These structures are in fact the discrete subgroups of the ambient Lie group. The geometrical symmetries of these structures can be computed in terms of the changes of generators of the discrete subgroup which preserve the discrete set of points. Here a classification of the symmetries for the discrete subgroups of a particular class of three-dimensional solvable Lie group is presented. It is a fact that there are only three mathematically distinct types of Lie groups which model uniform defective crystals, and the calculations given here complete the discussion of the symmetries of the corresponding discrete structures. We show that those symmetries corresponding to automorphisms of the discrete subgroups extend uniquely to symmetries of the ambient Lie group and we regard these symmetries as (restrictions of) elastic deformations of the continuous defective crystal. Other symmetries of the discrete structures are classified as `inelastic' symmetries.
\end{abstract}
\section{Introduction}
\label{intro}

In this paper it will be shown how the generalisation of symmetry properties of perfect solid crystals to crystals with certain uniform distributions of defects leads to a classification of the symmetries of discrete defective crystals as elastic or inelastic. We show that, in contrast with the perfect crystal case, some of the symmetries of a discrete defective crystal do not extend uniquely to a symmetry of the continuum model of the defective crystal. This allows us to classify the symmetries which don't extend as inelastic symmetries of the discrete defective crystal, while those which do extend uniquely are restrictions of elastic deformations of the continuous crystal and we call these symmetries elastic symmetries of the discrete crystal.

A starting point for the study of the mechanics of perfect solid crystals is to consider the geometrical symmetries of perfect lattices in $\mathbb{R}^3$:
\begin{equation}\label{eq:perflatt} L= \{ \bm x \in \mathbb{R}^3 \ : \ \bm x = n_a \bell_a,  \  n_a \in \mathbb{Z},\ a=1,2,3\}, \end{equation} where $\bell_1,\bell_2, \bell_3 \in \mathbb{R}^3$ are defining basis vectors and the summation convention operates on repeated indices. The perfect lattice $L$ defines a discrete set of points in $\mathbb{R}^3$ and can also be thought of as a discrete subgroup of the continuous Lie group $\mathbb{R}^3$ with addition as group composition. The discrete structures $L$ in $\mathbb{R}^3$ have geometrical symmetries $\bm \phi:L \to L$ given by \begin{equation}\label{eq:perfsymm}\bm \phi(\bell_a)=\gamma_{ab}\bell_b\qquad \mbox{where} \qquad \gamma = (\gamma_{ab}) \in GL_3(\mathbb{Z}).\end{equation} The symmetries of $L$ are bijective and preserve addition as well as the set of points in $\mathbb{R}^3$ defined by $L$. Moreover, every symmetry $\bm \phi$ of a perfect lattice $L$ extends uniquely to a bijection of $\mathbb{R}^3$.

Here we extend and generalise these properties of the symmetries of perfect crystals to a certain class of defective crystal where the distribution of defects is uniform. We use Davini's continuum model of defective crystals \cite{Davini}, in which the dislocation density tensor $S = (S_{ab})$, $a, b = 1, 2, 3,$ is defined by
\begin{equation}\label{eq:ddt}
    S_{ab}=\frac{\nabla \wedge \bm d_a \cdot \bm d_b}{ \bm d_1 \cdot \bm d_2 \wedge \bm d_3},
\end{equation}
where the fields $ \bm d_1(\cdot), \bm d_2(\cdot), \bm d_3(\cdot)$ are dual to the smooth \textit{lattice vector fields} $\bm \ell_1 (\cdot) , \bm \ell_2(\cdot),\bm \ell_3(\cdot)$, which represent the crystal geometry in a region $\Omega$. A crystal with a uniform distribution of defects has dislocation density tensor which is constant in space. (Note that for perfect crystals $S \equiv 0$.)

Suppose that $\{\bm\ell_a'(\cdot), \; a=1,2,3\}$ is a set of lattice vector fields which are elastically related to $\bm \ell_a(\cdot)$ in the sense that there exists a smooth invertible mapping $\bm u:\Omega\to \bm u (\Omega)\equiv \Omega'$ such that
\begin{equation}\label{eq:elast}
    \bm\ell_a'\left(\bm u(\bm x)\right)=\nabla \bm u(\bm x) \bm \ell_a(\bm x),\qquad \bm x\in \Omega,\;\  a=1,2,3.
\end{equation} We say that $\bm u(\cdot)$ is an elastic deformation. If $S'_{ab}$ is calculated via the analogue of \eqref{eq:ddt}, using fields dual to $\bm \ell'_a(\cdot),\; a=1,2,3$, then
\begin{equation}\label{1.3}
    S'_{ab} \left( \bm u\left(\bm x\right)\right)=S_{ab}(\bm x),\quad \bm x\in \Omega,\; a, b =1,2,3.
\end{equation}
Therefore, each component $S_{ab}(\cdot)$ of the dislocation density tensor is an `elastic' scalar invariant so that the value of the dislocation density is unchanged by elastic deformations of the crystal. Due to this elastic invariance, for a given dislocation density tensor $S$ there are infinitely many choices of (elastically related) sets of corresponding lattice vector fields.

It is a commonly held idea in the elasticity theory of perfect crystals that there is a continuum energy density $w$ which depends on the underlying perfect lattice $L$; that is $w=w(\{\bell_a\})$ where $\{\bell_a\}$ denotes the set of vectors $\{\bell_1, \bell_2, \bell_3\}$. If the basis vectors $\{\bell_a^\prime\}$ generate the same lattice $L$ (i.e. $\bell_a^\prime = \bm \phi(\bell_a)$ in  \eqref{eq:perfsymm}) then
\begin{equation} w(\{\bell_a\})=w(\{\bell_a^\prime\}).\end{equation} In other words the geometrical symmetries of $L$ correspond to the material symmetries of the energy density function $w$. Here we use a generalisation of this theory which accounts for the presence of a continuous distribution of defects in the crystal. It is assumed that the strain energy density per unit volume in such a crystal depends on the values of the lattice vector fields $\bell_1(\cdot)$, $\bell_2(\cdot)$, $\bell_3(\cdot)$ and dislocation density tensor $S(\cdot)$ at some point in $\Omega$. Thus
\begin{equation}w = w(\{\bell_a\}, S)\end{equation}
where $\{\bell_a\}$ denotes the set of vectors $\{\bell_1, \bell_2, \bell_3\}$ which are the values of the lattice vector fields at some point in $\Omega$, and $S$ denotes the dislocation density tensor evaluated at the same point. As we shall see in section \ref{sec:structure}, the arguments of the energy density function determine a structure $D$ which under certain conditions will be a discrete set of points. If another set of arguments, say $(\{\bell_a^\prime\}, S^\prime)$, determine the same discrete structure then it is assumed that
\begin{equation} w(\{\bell_a\}, S)= w(\{\bell_a^\prime\}, S^\prime).\end{equation} Hence for crystals with defects we are associating that the symmetries of the energy density function with the geometrical symmetries of the structure $D$. Therefore the structure $D$ is taken to be the defective crystal analogue of the perfect lattice $L$ underlying perfect crystals, and it is a central task to determine the geometrical symmetries of $D$.

Notice that the arguments $(\{\bell_a\}, S)$ of $w$ give no information regarding gradients of $S$, and we shall assume that they are zero. Therefore the dislocation density tensor $S$ is constant in space and the crystal has a uniform distribution of defects. It is this assumption that endows the crystal with a Lie group structure. Suppose that the vector fields $\bell_1(\cdot)$, $\bell_2(\cdot)$, $\bell_3(\cdot)$, defined here and henceforth on $\Omega\equiv \mathbb{R}^3$, give constant $S$. Then according to Pontryagin \cite{Pontryagin}, the system of partial differential equations
\begin{equation} \label{eq:intcond} \bell_a(\bm \psi(\bm x, \bm y))= \nabla_1\bm \psi(\bm x, \bm y)\bell_a(\bm x), \qquad a=1,2,3,\end{equation}
has a solution for the function $\bm \psi$, where  $\nabla_1\bm \psi(\cdot, \cdot)$ denotes the gradient of $\bm \psi$ with respect to its first argument. Moreover, the function $\bm \psi: \mathbb{R}^3 \times \mathbb{R}^3 \to \mathbb{R}^3$ can be taken to satisfy the properties required for it to be a Lie group composition function on $\mathbb{R}^3$, i.e.
\begin{eqnarray}  & \bm \psi(\mathbf{0}, \bm x) = \bm \psi(\bm x, \mathbf{0})= \bm x, \\ &\bm \psi(\bm x, \bm x^{-1})= \bm \psi(\bm x^{-1}, \bm x)= \mathbf{0}\\ & \bm \psi(\bm \psi(\bm x, \bm y), \bm z) = \bm \psi(\bm x, \bm\psi(\bm y, \bm z)),\end{eqnarray} where $\mathbf{0}$ is the group identity element and $\bm x^{-1}$ is the unique inverse of the element $\bm x$ \cite{Pontryagin,Parry03}. Here, the Lie group $G= (\mathbb{R}^3, \bm \psi)$ has underlying manifold $\mathbb{R}^3$ so that an element $\bm x \in G$ can be uniquely specified by $\bm x= x_i \be_i$ where $x_i \in \mathbb{R}$ and $\{ \be_1, \be_2, \be_3\}$ is a basis of $\mathbb{R}^3$. We will often use the alternative notation $\bm \psi(\bm x,\bm y) \equiv \bm x \bm y$. (Note that in the perfect crystal case where $S\equiv 0$, we can take $\bm \psi$ to be addition so that $G=(\mathbb{R}^3, +)$.)

Relation \eqref{eq:intcond} expresses the \textit{right invariance} of the fields $\{\bell_a(\cdot)\}$ with respect to the Lie group $G=(\mathbb{R}^3, \bm \psi)$. Suppose that the fields $\{\bell_a^\prime(\cdot)\}$ are elastically related to $\{\bell_a(\cdot)\}$ via $\bm u(\cdot)$ as in \eqref{eq:elast}. Then if $\bm \psi^\prime:\mathbb{R}^3 \times \mathbb{R}^3 \to \mathbb{R}^3$ is defined by
\[ \bm \psi^\prime(\bm r, \bm s) = \bm u(\bm \psi(\bm u^{-1}(\bm r), \bm u^{-1}(\bm s))), \] then
\[   \bm\ell_a'\left(\bm \psi^\prime(\bm r, \bm s)\right)=\nabla_1 \bm \psi^\prime(\bm r, \bm s) \bm \ell_a^\prime(\bm r),\qquad \bm r, \bm s \in \mathbb{R}^3,\;\  a=1,2,3.\] Hence the fields $\{\bell_a^\prime(\cdot)\}$ are right invariant with respect to the Lie group $G^\prime=(\mathbb{R}^3, \bm \psi^\prime)$ and since $\bm u$ is invertible, the groups $G$ and $G^\prime$ are isomorphic. Thus, elastically related crystal states have isomorphic corresponding Lie groups.

Recall that the dislocation density tensor $S$ is an elastic invariant so there is an infinite choice of elastically related lattice vector fields which have duals which satisfy \eqref{eq:ddt} for a given constant dislocation density. Therefore, there is also an infinite number of choices of isomorphic Lie groups $G= (\mathbb{R}^3, \bm \psi)$ corresponding to $S$. Hence a given constant $S$ determines up to Lie group isomorphism a Lie group $G$. In this paper we make a `canonical' choice of the Lie group $G$ to simplify the computations. We identify material points of the crystal with elements of the Lie group.

When $G$ has a uniform discrete subgroup $D$ (where $D \subset G$ is uniform if the left coset space $G/D$ is compact) the material points corresponding to the elements in $D$ have a minimum separation distance and form discrete geometrical structures which we take to the defective crystal analogue of the perfect lattice $L$. The requirement that $G/D$ be compact is a generalisation of the fact that in the perfect crystal case that $\mathbb{R}^3/L$ (the unit cell of the lattice $L$ with appropriate identification of boundary points) is compact.

According to Auslander, Green and Hahn \cite{Auslander} there are precisely three classes of non-Abelian, three dimensional Lie groups $G$ with uniform discrete subgroups. These are a certain class of nilpotent Lie groups and two non-isomorphic classes of solvable Lie groups. For each of these cases we are interested in the form of the geometrical structures corresponding to the discrete subgroups $D$ and the geometrical symmetries of these structures (i.e. the changes of generators of $D$ which preserve the points in the geometrical structure). These have already been determined in all three cases:
\begin{itemize}
\item When the Lie group $G$ is nilpotent (with corresponding Lie algebra with rational structure constants), Cermelli and Parry \cite{Cermelli06} have shown that the corresponding discrete subgroups give either a simple lattice or a 4-lattice (in Pitteri and Zanzotto's terminology \cite{Pitteri}) even though the composition function in $G$ is not additive. For such groups, Parry and Sigrist \cite{Parry11} construct explicitly all sets of generators of a given discrete subgroup. The formulae that connect different sets of generators generalise the perfect crystal case given by \eqref{eq:perfsymm}.
\item Auslander et al. \cite{Auslander} call the two classes of solvable groups $S_1$ and $S_2$. It has been shown by Parry and Nicks that in both cases the geometrical structures corresponding to the discrete subgroups of these solvable groups are simple lattices. The changes of generators preserving these structures were also determined (see \cite{Nicks1} for the $S_1$ case and \cite{Nicks2} for the $S_2$ case).
\end{itemize}

In this paper we focus on a property of the symmetries of perfect lattices which does not hold in the generalisation to the symmetries of discrete structures underlying crystals with uniform distributions of defects. Recall that the geometrical symmetries of a perfect lattice $L$ are bijections $\bm \phi:L \to L$ as in \eqref{eq:perfsymm} which preserve addition. Each of these symmetries extends uniquely to a bijection $\widetilde{\bm \phi}: \mathbb{R}^3\to \mathbb{R}^3$ defined by
\[ \widetilde{\bm \phi}(x_a \bell_a) = x_a(\widetilde{\bm \phi}(\bell_a))=x_a{\bm \phi}(\bell_a) , \qquad x_a \in \mathbb{R}, \ a=1,2,3.\]
Thus every symmetry of $L$ represents a (restriction of an) elastic deformation of the continuum perfect crystal.

This is not the case for crystals with constant $S\neq 0$ where the underlying Lie group, $G=(\mathbb{R}^3, \bm \psi)$, is solvable or nilpotent. In these cases there is a difference between the set of all geometrical symmetries of a discrete structure $D\subset G$ and the subset of these symmetries which preserve the group structure of $D$ and extend uniquely to elastic deformations of $\mathbb{R}^3$. This allows us to classify the symmetries of the discrete structures $D$ which preserve the group structure as elastic or inelastic depending on whether or not they are restrictions of elastic deformations of the continuum defective crystal. The observation that such a classification can be made is interesting because it indicates a possible link between the inelastic symmetries of the discrete crystal $D$ (which preserve the elastic invariant $S$ and the discrete structure) and observed inelastic processes in crystal behaviour such as slip in particular planes and directions determined by geometry.

Notice that we consider here only symmetries of discrete structures $D$ which additionally preserve the group structure. We do not discuss here symmetries where the discrete structure represents discrete subgroups of different Lie groups, isomorphic or not.

Our simplified task then is to identify which of the geometrical symmetries of discrete subgroups $D\subset G$ extend uniquely to elastic deformations of $\mathbb{R}^3$. This task breaks down into two stages. First we must determine which of the geometrical symmetries of $D$ preserve the group structure of $D$; that is which of the symmetries will extend to \textit{automorphisms} of $D$. Secondly we need to determine if these automorphisms of $D$ extend uniquely to automorphisms of the ambient Lie group $G$. For a geometrical symmetry of $D$ to be classified as an elastic symmetry it must extend to an automorphism of $D$ and that automorphism must extend uniquely to an automorphism of $G$, since these are requirements that must be satisfied in order that the geometrical symmetry is a restriction of an elastic deformation of the defective crystal.

In the cases where the structure $D$ is a discrete subgroup of a nilpotent Lie group or a solvable Lie group in the class $S_1$ such a classification of the geometrical symmetries of $D$ has been carried out. The automorphisms of the discrete subgroups $D$ have been computed (see \cite{Parry10} for the nilpotent case and \cite{Nicks3} for the $S_1$ case) and it has been observed that theorems of Mal'cev \cite{Malcev} and Gorbatsevich \cite{Gorbatsevich} guarantee that every automorphism of $D$ extends uniquely to an automorphism of the ambient Lie group $G$. In this paper we will complete the analysis by classifying the geometrical symmetries of discrete subgroups $D$ of solvable groups in the class $S_2$. In this case we must work a little harder since although it remains relatively straightforward to compute which of the geometrical symmetries of $D$ correspond to automorphisms of $D$, there is no analogue of the theorems of Mal'cev and Gorbatsevich for solvable groups of this class. Therefore we must determine directly whether or not automorphisms of $D$ extend uniquely to automorphisms of $S_2$. The difficulties arise for the $S_2$ class due to the fact that the exponential mapping from the corresponding Lie algebra $\mathfrak{s}_2$ to $S_2$ is not one-to-one.

We begin by recalling how to construct discrete structures $D$ corresponding to a particular set of arguments $(\{\bell_a\}, S)$ of the energy density function $w$. These are discrete subgroups of Lie groups $G$ and we will also recall elements of Lie group theory that will be required in this paper, including facts about Lie group isomorphisms. In section \ref{sec:solvable}, following Auslander et al. \cite{Auslander} and Nicks and Parry \cite{Nicks2} we introduce the group $S_2$ and the canonical group in the isomorphism class which we will work with. We also introduce the Lie algebra $\mathfrak{s}_2$ of the Lie group $S_2$ and calculate the automorphisms of $S_2$. In section \ref{sec:discrete} we discuss the discrete subgroups $D$ of $S_2$, recalling results from Nicks and Parry \cite{Nicks2} concerning their geometrical symmetries. We next compute the automorphisms of these discrete subgroups $D$ which amounts to determining the matrices $\chi \in GL_2(\mathbb{Z})$ which commute with a given matrix $\theta\in SL_2(\mathbb{Z})$ which is related to the dislocation density. This is a number theoretic problem studied by Baake and Roberts \cite{Baake} and here we summarize their results which are relevant to this work. Finally in section \ref{sec:extensions} we demonstrate explicitly that each of these automorphisms of $D$ extends uniquely to an automorphism of $S_2$.

\section{Elements of Lie group theory and discrete defective crystals} \label{sec:structure}

Suppose that we are given a set of arguments of an energy density function for a crystal with a uniform distribution of defects. That is, we are given $(\{\bell_a\}, S)$ where $S$ is some value of the dislocation density tensor and the (linearly independent) vectors $\bell_a$, $a=1,2,3$ are values of some lattice vector fields $\bell_a(\cdot)$ evaluated at some point, say $\mathbf{0}$, in $\mathbb{R}^3$ such that their duals satisfy \eqref{eq:ddt}. Furthermore (since we assume the crystal is uniform) the fields $\bell_a(\cdot)$ also satisfy \eqref{eq:intcond} for some group composition function $\bm \psi$ on $\mathbb{R}^3$. The following survey of facts about the Lie group $G= (\mathbb{R}^3, \bm \psi)$ follows that given in \cite{Nicks1}, \cite{Nicks2}, \cite{Nicks3}, \cite{Parry10}, \cite{Parry11} and is given here for completeness. The reader who is familiar with this background material may omit section \ref{sec:structure} and focus on the subsequent new material.

The Lie group $G=(\mathbb{R}^3, \bm \psi)$ has corresponding Lie algebra $\mathfrak{g}$ which is the vector space $\mathbb{R}^3$ with the Lie bracket operation $\left[ \cdot, \cdot\right]:\mathbb{R}^3 \times \mathbb{R}^3 \to \mathbb{R}^3$ given by
\begin{equation} \label{eq:bracket} \left[ \bm x, \bm y\right]= C_{ijk} x_j y_k \bm e_i, \qquad \bm x, \bm y \in \mathbb{R}^3,\end{equation}
with respect to some basis $\{\bm e_1, \bm e_2, \bm e_3\}$ of $\mathbb{R}^3$. Here, $C_{ijk}$ are the structure constants of the Lie algebra and are related to the Lie group composition function $\bm \psi$ via
\begin{equation}\label{eq:strucconsts} C_{ijk}=\frac{\partial^2\psi_i}{\partial x_j\partial y_k}(\bm 0, \bm 0) - \frac{\partial^2\psi_i}{\partial x_k\partial y_j}(\bm 0, \bm 0),\end{equation} where $\bm \psi = \psi_i(\bm x, \bm y)\bm e_i$. The connection between the dislocation density tensor ${S}$, defined via \eqref{eq:ddt}, and the structure constants is
\begin{equation} \label{eq:ddtstrucconsts} C_{ijk} \ell_{rj}(\bm 0) \ell_{sk}(\bm 0)= \epsilon_{prs}{S}_{kp}\ell_{ki}(\bm 0),\end{equation} where $\epsilon_{prs}$ is the permutation symbol and $\bm \ell_r(\bm 0) = \ell_{rj}(\bm 0)\bm e_j$, see Elzanowski and Parry \cite{Elzanowski}.

In this paper we shall be concerned with the automorphisms of Lie groups $G$ which are of course isomorphisms of the Lie group to itself (preserving the group composition function). These are related to the automorphisms of the corresponding Lie algebra $\mathfrak{g}$. Let $\bm{\mathfrak{g}}$ and $\bm{\mathfrak{g}}'$  be Lie algebras with Lie brackets $[\cdot,\cdot]_{\bm{\scriptstyle{\mathfrak{ g}}}},\;[\cdot,\cdot]_{\bm{\scriptstyle{\mathfrak{ g'}}}}$ respectively.  A Lie algebra isomorphism is an invertible linear transformation $L:\bm{\mathfrak{g}}\to \bm{\mathfrak{g}}'$ which satisfies
\begin{equation}\label{eq:algebraiso}
    [L\bm x,L\bm y]_{\bm{\scriptstyle{\mathfrak{ g'}}}}=L[\bm x,\bm y]_{\bm{\scriptstyle{\mathfrak{ g}}}},\quad \bm x, \bm y\in \bm{\mathfrak g}.
\end{equation} If $C^{\bm{\scriptstyle{\mathfrak{ g}}}}_{ijk}, C^{\bm{\scriptstyle{\mathfrak{ g'}}}}_{ijk}$ are the structure constants for $\bm{\mathfrak{g}}, \bm {\mathfrak g}'$ respectively, then \eqref{eq:bracket} implies that
\begin{equation}\label{eq:algebraconditions}
  C^{\bm{\scriptstyle{\mathfrak{ g'}}}}_{ijk} L_{jp} L_{kq}= L_{ir} C^{\bm{\scriptstyle{\mathfrak{ g}}}}_{rpq},
\end{equation}
where $L\bm e_i=L_{ji} \bm e_j, \; i=1,2,3$. Let $G=(\mathbb{R}^3, \bm \psi_G)$ and $G^\prime=(\mathbb{R}^3, \bm \psi_{G^\prime})$ be Lie groups with corresponding Lie algebras $\bm{\mathfrak{g}}$ and $\bm{\mathfrak{g}}'$ respectively. A smooth invertible mapping $\bm u:G \to G^\prime$ is a Lie group isomorphism if
\begin{equation} \bm \psi_{G^\prime}(\bm u(\bm x), \bm u(\bm y))= \bm u(\bm \psi_G(\bm x, \bm y)), \qquad \bm x, \bm y \in G.\end{equation}
It is a fact that if $\bm u:G\to G^\prime$ is a Lie group isomorphism then $\nabla \bm u(\bm 0)\equiv L$ is a Lie algebra isomorphism from $\bm{\mathfrak{g}}$ to $\bm{\mathfrak{g}}'$.  Conversely, if an invertible linear transformation $L$ satisfies \eqref{eq:algebraiso}, then it is a major result of Lie theory that there exists a unique Lie group isomorphism $\bm u$ such that $\nabla \bm u (\bm 0)=L$ (see \cite{Varadarajan}).

Let $\nu_1, \nu_2, \nu_3$ be given real numbers and define the right invariant vector field $\bm \nu(\cdot) = \nu_a \bell_a(\cdot)$. Define the integral curve of $\bm \nu(\cdot)$ through $\bm x_0$ to be the solution $\{ \bm x(t)\ : t\in \mathbb{R}\}$ of the differential equation
$\dot{\bm x} (t)= \nu_a \bell_a(\bm x (t))$, $\bm x(0)=\bm x_0$. Note that $\bm \nu:=\bm \nu(\bm 0)$ determines the field $\bm \nu (\bm x)$ by the right invariance of $\bm \nu(\cdot)$. One can then define the mapping $\exp(\bm \nu):G \to G$ by
\begin{equation}\label{eq:exponentialmapping}
\exp(\bm \nu)(\bm x_0)= \bm x(1),
\end{equation}
and the group element ${\rm e}^{(\bm \scriptstyle{\nu})}$ by
\begin{equation}\label{eq:exponentialelement}
{\rm e}^{(\bm{\scriptstyle{\nu}})}= \exp(\bm \nu)(\bm 0).
\end{equation} Also, note that ${\rm e}^{(\cdot)}:\mathfrak{g} \to G$ is called the exponential mapping of the Lie algebra to the Lie group. It is standard result of Lie group theory that
\begin{equation}\label{eq:flowandmultiplication}
\exp(\bm \nu)(\bm x) = \bm \psi({\rm e}^{(\bm{\scriptstyle{\nu}})}, \bm x),
\end{equation}
and this states that the flow along the integral curves of the lattice vector fields corresponds to group multiplication by the group element ${\rm e}^{(\bm{\scriptstyle{\nu}})}$. In the case of perfect crystals, choosing $\bell_a(\cdot)\equiv \bell_a(\bm 0)\equiv \be_a$ for a basis $\{ \be_1, \be_2, \be_3\}$ of $\mathbb{R}^3$, iterating the flow along the lattice vector fields (which in this case is just translation by $\be_1, \be_2, \be_3$) produces a perfect lattice. In the case of $G=(\mathbb{R}^3, \bm \psi)$ the analogue of the perfect lattice is the set of points (or group elements) produced by iterating the flow (from $t=0$ to $t=1$) along the lattice vector fields, starting at the origin. By \eqref{eq:exponentialmapping}--\eqref{eq:flowandmultiplication} one obtains the subgroup of $G$ that is generated by the group elements ${\rm e}^{(\bm{\scriptstyle{e}}_1)}$, ${\rm e}^{(\bm{\scriptstyle{e}}_2)}$, ${\rm e}^{(\bm{\scriptstyle{e}}_3)}$ where $\bm e_a= \bell_a(\bm 0)$, $a=1,2,3$.

\begin{figure}[h]
  \centering
  \includegraphics[width=5cm]{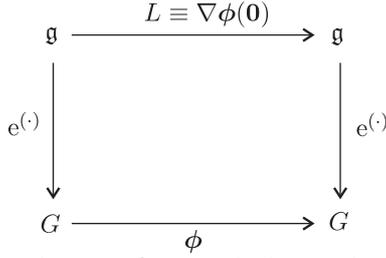}\\
  \caption{Commutative diagram for Lie algebra and Lie group automorphisms}\label{fig:commuting}
\end{figure}

We will be interested in the automorphisms of both $G$ and its subgroup $D$ generated by ${\rm e}^{(\bm{\scriptstyle{e}}_1)}$, ${\rm e}^{(\bm{\scriptstyle{e}}_2)}$, ${\rm e}^{(\bm{\scriptstyle{e}}_3)}$ whenever this is a uniform discrete subgroup. The method we will use to compute the automorphisms of $G$ makes use of the fact that the diagram in Figure \ref{fig:commuting} commutes so that
\begin{equation}\label{eq:commuting}
\bm \phi({\rm e}^{(\bm{\scriptstyle{\nu}})}) = {\rm e}^{({\scriptstyle{\nabla}} \bm{\scriptstyle{\phi}}\bm{\scriptstyle{(0)}}\bm{\scriptstyle{\nu}})}, \qquad \bm \nu \in \mathfrak{g},
\end{equation} and also the fact that the automorphisms of the corresponding Lie algebra $\mathfrak{g}$ can be computed using the fact that they must satisfy \eqref{eq:algebraconditions} with $\mathfrak{g}^\prime= \mathfrak{g}$.

\section{Solvable Lie groups and their automorphisms} \label{sec:solvable}

Recall that in this paper we shall be completing the classification of symmetries of discrete structures associated with crystals with uniform distributions of defects. As we have seen in section \ref{sec:structure}, the discrete structures are uniform discrete subgroups $D$ of three-dimensional Lie groups $G=(\mathbb{R}^3, \bm \psi)$. According to Auslander et al \cite{Auslander}, there are only three classes of non-abelian, connected, simply connected, three-dimensional Lie groups $G=(\mathbb{R}^3, \bm \psi)$ which have such uniform discrete subgroups. These are a class of nilpotent Lie group and two non-isomorphic classes of solvable Lie group which they call $S_1$ and $S_2$. The symmetries of the discrete subgroups of the nilpotent Lie groups and the solvable groups in the class $S_1$ have already been computed and classified. Here we complete the analysis by considering the $S_2$ case. We begin with the relevant definitions.

\subsection{Solvable Lie groups}
Let $\mathfrak{g}$ be a Lie algebra with corresponding connected Lie group $G$. Define the following sequence of subalgebras:
\[\mathfrak{g}_1= \mathfrak{g}, \quad \mathfrak{g}_2= \left[ \mathfrak{g}_{1}, \mathfrak{g}_{1}\right], \quad \ldots, \quad\mathfrak{g}_k= \left[ \mathfrak{g}_{k-1}, \mathfrak{g}_{k-1}\right].\] The Lie algebra $\mathfrak{g}$ is solvable if $\mathfrak{g}_k= \bm 0$ for some integer $k$.
Let $(\bm x, \bm y) = \bm x^{-1}\bm y^{-1} \bm x \bm y$ denote the commutator of $\bm x, \bm y \in G$ where group multiplication is represented as juxtaposition. Then let $(G,G)$ denote the commutator (or derived) subgroup of $G$ generated by all commutators of elements of $G$. If one defines
\[ G_1=G, \quad G_2= (G_1,G_1), \quad G_3=(G_2, G_2), \quad \ldots, \quad G_k=(G_{k-1}, G_{k-1}),\] the $G$ is solvable if $G_k=\bm 0$ for some integer $k$. The Lie algebra of $G_k$ is $\mathfrak{g}_k$ and $G$ is solvable if and only if $\mathfrak{g}$ is solvable.

In solvable groups of dimension three we have $G_3=\bm 0$ so that all commutators of elements of $G$ commute with each other. Furthermore, it can be shown that there are basis vectors $\bm f_1$, $\bm f_2$, $\bm f_3$ of $\mathbb{R}^3$ such that
\begin{equation} \label{eq:algbasis}
\left[ \bm f_1, \bm f_2 \right] = \bm 0, \qquad  \left[ \bm f_1, \bm f_3 \right] = \alpha \bm f_1 + \beta \bm f_2, \qquad \left[ \bm f_2, \bm f_3 \right] =\gamma  \bm f_1 + \delta \bm f_2,
\end{equation}
where $\alpha, \beta, \gamma, \delta \in \mathbb{R}$ and $\alpha\delta-\beta\gamma\neq 0$.

\subsection{The solvable Lie group $S_2$}

We shall be concerned with the three-dimensional solvable Lie group $S_2$ which has corresponding Lie algebra which we shall denote $\mathfrak{s}_2$. We now define this group and give the form of corresponding dislocation density tensor. Further details regarding derivation of facts about this group or the related group $S_1$ can be found in Nicks and Parry \cite{Nicks2} and Nicks and Parry \cite{Nicks1} respectively. We identify group elements with points $\bm x \in \mathbb{R}^3$, representing them as $\bm x= x_i \bm e_i$ with respect to some basis $\{\bm e_1, \bm e_2, \bm e_3\}$ of $\mathbb{R}^3$. Auslander et al \cite{Auslander} choose to represent the elements as $4 \times 4$ matrices (still parameterised by $x_1, x_2, x_3$) and these matrix representations form an isomorphic group $S_m$ where the matrix representation of $\bm x \in S_2$ is $r_m(\bm x) \in S_m$, defined by
\begin{equation} \label{eq:elements}
r_m(\bm x)\equiv \left( \begin{array}{cc}
\\[-0.3cm]
\multicolumn{2}{c|}{\phi(x_3)}\\[0.3cm]
\hline
0&0\\
0&0\end{array}
\begin{array}{|cc}
0&x_1\\0&x_2\\\hline
1&x_3\\
0&1\end{array}
\right), \quad\bm x\equiv \left(\begin{array}{c} x_1\\x_2\\x_3\end{array}\right) \in \mathbb{R}^3, \quad \phi(x_3)=\left( \begin{array}{cc}a(x_3)&b(x_3)\\ c(x_3)&d(x_3)\end{array}\right).
\end{equation}
In \eqref{eq:elements}, $\phi(x_3) \in SL_2 (\mathbb{R}),\  \phi(1) \in SL_2(\mathbb{Z})$ and $\left\{\phi(x_3): x_3 \in \mathbb{R}\right\}$ is a one parameter subgroup of the unimodular group. This implies that
\begin{equation}\label{eq:oneparameter}
\phi(x)\phi(y)=\phi(x+y), \qquad x,y, \in \mathbb{R},
\end{equation}
and hence $\phi(0)=\mathbb{I}_2$, the $2 \times 2$ identity matrix. The one parameter subgroups of $SL_2(\mathbb{R})$ which have $\phi(1) \in SL_2(\mathbb{Z})$ fall into two classes depending on the eigenvalues of $\phi(1)$. Let us define
\begin{equation}\label{eq:theta}
    \phi(1)\equiv \theta = \left( \begin{array}{cc}a(1)&b(1)\\ c(1)&d(1)\end{array}\right) = \left( \begin{array}{cc}a&b\\ c&d\end{array}\right), \quad a,b,c,d \in \mathbb{Z}, \ \ ad-bc=1.
 \end{equation}
 The eigenvalues, $\lambda$, of $\theta$ are roots of the characteristic polynomial $P(\lambda)= \lambda^2 - \mbox{tr}(\theta)\lambda+1$,
 and are therefore given by $\lambda, 1/\lambda$ where $\lambda = \tfrac{1}{2}(\mbox{tr}(\theta) +\sqrt{\mbox{tr}(\theta)^2-4})$.
 If $\mbox{tr}(\theta)\in \{-2,-1,0,1\}$ then the eigenvalues of $\theta$ are a complex conjugate pair and the group of matrices of the form $r_m(\bm x)$ is isomorphic to $S_2$. Moreover, if $\mbox{tr}(\theta)=-2$ then $\theta\equiv -\mathbb{I}_2$. If $\mbox{tr}(\theta)>2$ then the eigenvalues of $\theta$ are positive, real and distinct and the group of matrices of the form $r_m(\bm x)$ is isomorphic to the class of solvable group $S_1$ which has been considered previously (see Nicks and Parry \cite{Nicks3}). If $\mbox{tr}(\theta)=2$ then the group of matrices of the form $r_m(\bm x)$ is isomorphic to a nilpotent Lie group. For other values of $\mbox{tr}(\theta)$, $\theta$ cannot lie on a one parameter subgroup of $SL_2(\mathbb{R})$ (see Auslander et al. \cite{Auslander}).

 Differentiating \eqref{eq:oneparameter} with respect to $y$ and evaluating at $y=0$, and also doing the same for $x$, we see that the one parameter subgroup $\phi(x)$ of $SL_2(\mathbb{R})$ satisfies
 \[ \phi^\prime(x) = \phi(x)\phi^\prime(0)= \phi^\prime(0)\phi(x),\] where ${}^\prime$ denotes $\frac{d}{dx}$. We also define
 \[ \phi^\prime(0) = \mathcal{A} = \left( \begin{array}{cc}a'(0)&b'(0)\\ c'(0)&d'(0)\end{array}\right),\] so that we have
 \[ \phi(x) = {\rm e}^{\mathcal{A}x} = \sum_{j=0}^\infty \mathcal{A}^j \frac{x^j}{j!}.\] Since $\phi(x) \in SL_2(\mathbb{R})$, we have $a(x)d(x)-b(x)c(x)=1$ and differentiating this relation with respect to $x$ and setting $x=0$ we see that $\mbox{tr}(\mathcal{A})=0$ since $\phi(0)= \mathbb{I}_2$. Hence $\mathcal{A}^2 =- \det (\mathcal{A}) \mathbb{I} _2$ and any matrix satisfying this condition has matrix exponential satisfying
 \begin{equation}\label{2.10}
    {\rm e}^{\mathcal{A}}=\left\{ \begin{array}{ll}
    (\cosh k) \mathbb{I}_2+\left( \frac{\sinh k}{k}\right) \mathcal{A},\quad&\text{ if } \det (\mathcal{A})<0,\quad k\equiv \sqrt{-\det(\mathcal{A})};\\[0.3cm]
    (\cos k) \mathbb{I}_2+\left( \frac{\sin k}{k}\right) \mathcal{A},\quad&\text{ if } \det (\mathcal{A})>0,\quad k\equiv \sqrt{\det(\mathcal{A})};\\[0.3cm]
      \mathbb{I}_2+\mathcal{A}, \quad&\text{ if } \det (\mathcal{A})=0.
    \end{array}\right.
\end{equation}

Since $\mbox{tr}(\mathcal{A})=0$ we have
\begin{equation}
    a + d =\mbox{tr} \; {\rm e}^{\mathcal{A}}= \left\{ \begin{array}{ll} 2\cosh k, \quad &\text{ if } \det (\mathcal{A})<0;\\[0.3cm]
    2\cos k, \quad &\text{ if } \det (\mathcal{A})>0;\\[0.3cm]
    2, \quad &\text{ if } \det (\mathcal{A})=0,\end{array}\right.
\end{equation} where $k= \sqrt{|\det(\mathcal{A})|}$. Recall that we are interested in the case where $S_m$ is isomorphic to $S_2$ where
$a+d \in \{-2,-1,0,1\}$. In that case $\det(\mathcal{A})>0$ and
\begin{equation}\phi(x) = {\rm e}^{\mathcal{A}x} =  (\cos kx) \mathbb{I}_2+\left(\frac{\sin kx}{k}\right) \mathcal{A}, \end{equation} where\begin{equation} \label{eq:k} k = \left\{\begin{array}{lll}
\pi n & \  n=\pm 1  \mod 2 & \ \mbox{when}\  a+d=-2\\
\frac{2\pi n}{3}& \  n=\pm 1 \mod 3 & \ \mbox{when}\  a+d=-1\\
\frac{\pi n}{2}& \  n=\pm 1\mod 4 &\  \mbox{when}\  a+d=0\\
\frac{\pi n}{3}&\  n=\pm 1 \mod 6 & \ \mbox{when}\  a+d=1.\\
\end{array} \right.\end{equation}

When $a+d=-2$ so that $\theta=-\mathbb{I}_2$, $\mathcal{A}$ is any traceless $2 \times 2$ matrix with determinant $k^2=n^2\pi^2$, $n$ an odd integer. If $a+d \in \{-1,0,1\}$ then
\[ \theta= \left( \begin{array}{cc}a&b\\ c&d\end{array}\right) = \frac{1}{2} (a+d) \mathbb{I}_2 + \left(\frac{\sin k}{k}\right) \mathcal{A}\] which implies that
\begin{equation}\label{eq:A} \mathcal{A} = \left(\frac{k}{\sin k}\right)\left( \begin{array}{cc}\tfrac{1}{2}(a-d) &b\\ c&-\tfrac{1}{2}(a-d)\end{array}\right).\end{equation} Note that in the cases where $a+d \in \{-1,0,1\}$ the integers $b$ and $c$ in $\mathcal{A}$ and $\theta$ are nonzero, since if either were zero then it must be the case that $ad=1$ and hence $a=d=\pm 1$, implying that $a+d=\pm 2$.

Noting that the mapping $r_m: S_2 \to S_m$ given by \eqref{eq:elements} is one to one and that matrix multiplication is the group composition function in the matrix group $S_m$, it follows that the group composition function $\bm \psi$ in $S_2$ satisfies
\[ r_m(\bm \psi(\bm x, \bm y)) =  r_m(\bm x) r_m(\bm y),\] and hence
\begin{equation}\label{eq:groupmult}
\bm \psi(\bm x, \bm y) = \bm x + (a(x_3)y_1 + b(x_3)y_2)\bm e_1 + (c(x_3)y_1 + d(x_3)y_2)\bm e_2 + y_3 \bm e_3.
\end{equation}

For a given a group composition function $\bm \psi$ it is easy to see that $\bell_a(\bm x) = \nabla_1 \bm \psi(\bm 0, \bm x)\bm e_a$, $a=1,2,3$, is a set of lattice vector fields which is right invariant with respect to the group $(\mathbb{R}^3, \bm \psi)$; that is \eqref{eq:intcond} is satisfied where $\bell_a(\bm 0)=\bm e_a$. For our group composition function \eqref{eq:groupmult} we have
\[ \bell_1(\bm x)= \bm e_1, \ \  \bell_2(\bm x)= \bm e_2, \ \  \bell_3(\bm x)= (a^\prime(0)x_1 + b^\prime(0)x_2)\bm e_1 + (c^\prime(0)x_1 - a^\prime(0)x_2)\bm e_2+ \bm e_3,\] recalling that $d^\prime(0)=-a^\prime(0)$. Using the duals of these vector fields we then compute using \eqref{eq:ddt} that the components of the dislocation density tensor are
\begin{equation} \label{eq:S2ddt} S= \left( \begin{array}{ccc}-b^\prime(0)& a^\prime(0) & 0\\ a^\prime(0)&c^\prime(0)& 0\\ 0 & 0 & 0 \end{array}\right). \end{equation} In particular the dislocation density tensor is rank 2, symmetric and uniquely determined by the matrix $\mathcal{A}$. Also note that the correspondence between the $2 \times 2$ matrices $\mathcal{A}$ and $\theta= {\rm e}^\mathcal{A}$ is infinitely many to one since $\theta$ determines $\mbox{tr}(\theta)=a+d=2\cos k$ where $k$ may take a countable infinity of values.

\subsection{The solvable Lie algebra $\mathfrak{s}_2$ and its automorphisms}

From the group composition function $\bm \psi$ in $S_2$ given by \eqref{eq:groupmult} one can calculate using \eqref{eq:bracket} that the Lie bracket on $\mathfrak{s}_2$ (the Lie algebra of $S_2$) is given by
\begin{equation}\label{eq:S2brac} \left[ \bm x, \bm y\right]= (a^\prime(0) \bm x \wedge \bm y \cdot \bm e_2 - b^\prime(0)\bm x \wedge \bm y \cdot \bm e_1) \bm e_1 +(c^\prime(0) \bm x \wedge \bm y \cdot \bm e_2 + a^\prime(0)\bm x \wedge \bm y \cdot\bm e_1) \bm e_2 , \end{equation} for $\bm x= x_i\bm e_i$, $\bm y= y_i \bm e_i \in \mathbb{R}^3$. In particular
\[ \left[ \bm e_1, \bm e_2\right] = \bm 0, \ \ \left[ \bm e_1, \bm e_3\right]= -a^\prime(0)\bm e_1 -c^\prime(0)\bm e_2, \ \ \left[ \bm e_2, \bm e_3\right]= -b^\prime(0)\bm e_1 +a^\prime(0)\bm e_2,\] so that it is clear from \eqref{eq:algbasis} that $\mathfrak{s}_2$ is solvable since $\det(\mathcal{A})\neq 0$.

We now want to compute the automorphisms of this Lie algebra. It turns out that these computations are much simplified if we make a change of basis in the Lie algebra. Suppose that we are given a particular value of $\theta$ with $\mbox{tr}(\theta) \in \{-1,0,1\}$ or $\theta=-\mathbb{I}_2$ and make a choice of corresponding $k$ (c.f. \eqref{eq:k}). Consider the basis $\{\bm f_1, \bm f_2, \bm f_3\}$ where $\bm f_{i}= M_{ij}\bm e_j$ for
\begin{equation}\label{eq:M} M= \left( \begin{array}{ccc}-b^\prime(0)& a^\prime(0)+k & 0\\ -b^\prime(0)& a^\prime(0)-k& 0\\ 0 & 0 & 1\end{array}\right).\end{equation}
Note that $b^\prime(0)\neq 0$ for $\mbox{tr}(\theta) \in \{-1,0,1\}$ by the remark following \eqref{eq:A} and if $b^\prime(0)=0$ when $\mbox{tr}(\theta) =-2$ then $\det(\mathcal{A})=-(a^\prime(0))^2<0$ since $a^\prime(0) \in \mathbb{R}$ but this contradicts the fact that $\det(\mathcal{A})=k^2>0$. Thus the change of basis matrix $M$ is invertible since it has $\det(M)=2b^\prime(0)k\neq 0$.

The basis $\{\bm f_1, \bm f_2, \bm f_3\}$ satisfies
\[ \left[ \bm f_1, \bm f_2\right] = \bm 0, \ \ \left[ \bm f_1, \bm f_3\right]= k \bm f_2, \ \ \left[ \bm f_2, \bm f_3\right]= -k \bm f_1.\] Furthermore, with respect to this basis the structure constants of $\mathfrak{s}_2$ are given by
\begin{equation}\label{eq:s2strucconsts} C_{ijk} = k(\delta_{3j}\epsilon_{3ik}- \delta_{3k}\epsilon_{3ij}),\end{equation} and using \eqref{eq:algebraconditions} with $\mathfrak{g}=\mathfrak{g}^\prime = \mathfrak{s}_2$ one computes that a linear transformation $L$ is an automorphism of $\mathfrak{s}_2$ with respect to the basis $\{\bm f_1, \bm f_2, \bm f_3\}$ if $L$ has the form (c.f. \cite{Ha})
\begin{equation}\label{eq:s2auto} L= \left( \begin{array}{rrr}0& 1 & 0\\ 1& 0& 0\\ 0 & 0 & -1 \end{array}\right)^{\epsilon}\left( \begin{array}{rrr}\alpha& \beta & \gamma\\ -\beta& \alpha& \delta\\ 0 & 0 & 1 \end{array}\right),  \end{equation} where $\epsilon \in \{0,1\}$, $\alpha,\beta,\gamma, \delta \in \mathbb{R}$ such that $\alpha^2 +\beta^2\neq 0$.
With respect to the basis $\{\bm e_1, \bm e_2, \bm e_3\}$, the automorphisms of $\mathfrak{s}_2$ are given by $M^TLM^{-T}$ where $T$ denotes transpose.

\subsection{Automorphisms of $S_2$}\label{sec:S2auto}

Here we discuss how to compute the group automorphisms of $S_2$. As previously discussed we shall use relation \eqref{eq:commuting} to compute these using our knowledge of the Lie algebra automorphisms of $\mathfrak{s}_2$. This computation is not as straight forward as it may appear at first glance, due to the fact that the exponential mapping ${\rm e}^{(\cdot)}:\mathfrak{s}_2 \to S_2$ is not surjective. Again we work with respect to the basis $\{\bm f_1, \bm f_2, \bm f_3\}$ for ease of computation. We begin by giving details of the required functions and mappings with respect to the basis $\{\bm f_1, \bm f_2, \bm f_3\}$.

When changing basis from $\{ \bm e_1, \bm e_2, \bm e_3\}$ to $\{\bm f_1, \bm f_2, \bm f_3\}$, the matrix $\mathcal{A}$ changes to $\mathcal{B}=  \bigl(\begin{smallmatrix}0&k\\ -k&0\end{smallmatrix} \bigr)$ and, with respect to the basis $\{\bm f_1, \bm f_2, \bm f_3\}$,
\begin{equation}\label{eq:compf} \phi(u)= {\rm e}^{\mathcal{B}u} = \left( \begin{array}{rr}\cos ku &\sin ku \\ -\sin ku& \cos ku \end{array}\right).\end{equation}  Thus with respect to the basis $\{\bm f_1, \bm f_2, \bm f_3\}$ the Lie group composition function in $S_2$ is given by
\[
\bm \psi(\bm u, \bm v) = \bm u + (v_1 \cos ku_3 + v_2 \sin ku_3)\bm f_1 + (-v_1\sin ku_3 + v_2 \cos ku_3)\bm f_2 + v_3 \bm f_3
\]
where $\bm u= u_i \bm f_i$, $\bm v= v_i \bm f_i$. Computing the lattice vector fields $\nabla_1 \bm \psi(\bm 0, \bm u)\bm f_a$ one can then use \eqref{eq:exponentialelement} to find that the exponential mapping ${\rm e}^{(\cdot)}: \mathfrak{s}_2 \to S_2$ with respect to the basis $\{\bm f_1, \bm f_2, \bm f_3\}$ is given by
\begin{equation} \label{eq:S2exp}
{\rm e}^{(\bm \scriptsize{u})} = \left( \begin{array}{c} F(\mathcal{B}u_3)\left(\begin{array}{c}u_1 \\u_2\end{array}\right) \\ u_3\end{array}\right) \qquad \mbox{where}\ \bm u= u_i \bm f_i = (u_1, u_2, u_3)^T,
\end{equation}
and for $u_3\neq 0$
\begin{eqnarray}
F(\mathcal{B}u_3) &=& \sum_{j=0}^\infty \frac{(\mathcal{B} u_3)^j}{(j+1)!} = \left( \frac{\sin ku_3}{k u_3}\right) \mathbb{I}_2 + \left( \frac{1-\cos ku_3 }{(k u_3)^2}\right)\mathcal{B}u_3\nonumber \\ &=& \frac{1}{ku_3} \left( \begin{array}{cc} \sin ku_3 & 1-\cos ku_3 \\ -(1-\cos ku_3) & \sin ku_3 \end{array}\right).\label{eq:F}\end{eqnarray}
For $u_3=0$, $F(\mathcal{B}u_3) = \mathbb{I}_2$.

Note that if $ku_3= 2\pi n$ for some $n \in \mathbb{Z}\backslash\{0\}$ then $F(\mathcal{B}u_3) = \bm 0$, the $2 \times 2$ zero matrix, and ${\rm e}^{(\bm \scriptsize{u})} = (0,0,u_3)^T$ for any values of $u_1$ and $u_2 \in \mathbb{R}$. Therefore ${\rm e}^{(\cdot)}: \mathfrak{s}_2 \to S_2$ is not a homeomorphism - it is not surjective because it is not possible to write every element $\bm v \in S_2$ as $\bm v={\rm e}^{(\bm \scriptsize{u})}$ for some $\bm u \in \mathfrak{s}_2$. Hence the exponential mapping ${\rm e}^{(\cdot)}: \mathfrak{s}_2 \to S_2$ does not have a well defined inverse.

However, one can verify that any $\bm v = v_i \bm f_i \in S_2$ may be written as the group composition of two exponentials:
\begin{equation}\label{eq:twoexp} \bm v = \bm \psi({\rm e}^{(\bm \scriptsize{s})}, {\rm e}^{(\bm \scriptsize{t})}) \qquad \mbox{where} \  \bm s = v_1 \bm f_1 + v_2 \bm f_2, \ \ \bm t = v_3 \bm f_3.\end{equation}
Recall that there is a one to one correspondence between the Lie algebra automorphisms $L:\mathfrak{s}_2 \to \mathfrak{s}_2$ and the Lie group automorphisms $\bm \phi:S_2 \to S_2$ given by $L\equiv \nabla \bm \phi(\bm 0)$. Also the Lie group and algebra automorphisms satisfy \eqref{eq:commuting}. We now use this relation to compute the Lie group automorphisms $\bm \phi:S_2 \to S_2$. Suppose that $L:\mathfrak{s}_2 \to \mathfrak{s}_2$ is a Lie algebra automorphism and hence has the form \eqref{eq:s2auto}, and let $\bm v \in S_2$. Then $\bm v$ can be written as in \eqref{eq:twoexp} so that Lie group automorphisms $\bm \phi:S_2 \to S_2$ satisfy
\[ \bm \phi (\bm v) = \bm \phi(\bm \psi({\rm e}^{(\bm \scriptsize{s})}, {\rm e}^{(\bm \scriptsize{t})})) = \bm \psi(\bm \phi({\rm e}^{(\bm \scriptsize{s})}), \bm \phi({\rm e}^{(\bm \scriptsize{t})})) = \bm \psi({\rm e}^{(\bm \scriptsize{Ls})}, {\rm e}^{(\bm \scriptsize{Lt})}),\]
where $\bm s$ and $\bm t$ are as in \eqref{eq:twoexp}. Hence ${\rm e}^{(\bm \scriptsize{Ls})}= L\bm s$ and
\[ {\rm e}^{(\bm \scriptsize{Lt})} =  \left( \begin{array}{c} F(\mathcal{B}\xi v_3)W(\epsilon) \left(\begin{array}{c}\gamma \\ \delta\end{array}\right)v_3 \\ \xi v_3\end{array}\right), \quad \mbox{where} \ W(\epsilon)= \left(\begin{array}{cc}0 & 1 \\1 & 0\end{array}\right)^\epsilon,\ \xi=(-1)^\epsilon, \] so that automorphisms of $S_2$ with respect to the basis $\{\bm f_1, \bm f_2, \bm f_3\}$ are given by
\begin{eqnarray}\label{eq:S2auto}\bm \phi (\bm v) &=&  \left( \begin{array}{c} W(\epsilon) \left(\begin{array}{cc}\alpha & \beta \\-\beta & \alpha\end{array}\right) \left(\begin{array}{c}v_1 \\v_2\end{array}\right) + F(\mathcal{B}\xi v_3)W(\epsilon) \left(\begin{array}{c}\gamma \\ \delta\end{array}\right) v_3    \\ \xi v_3\end{array}\right)\\
&=& \nonumber \left\{ \begin{array}{ll} \left(\begin{array}{c}\alpha v_1 + \beta v_2 + \tfrac{\gamma}{k} \sin kv_3 + \tfrac{\delta}{k}(1-\cos kv_3) \\ -\beta v_1 + \alpha v_2 - \tfrac{\gamma}{k}(1-\cos kv_3)  + \tfrac{\delta}{k}\sin kv_3\\ v_3\end{array}\right) & \mbox{ when } \epsilon=0,  \\ \left(\begin{array}{c} -\beta v_1 + \alpha v_2 - \tfrac{\gamma}{k}(1-\cos kv_3)  + \tfrac{\delta}{k}\sin kv_3\\ \alpha v_1 + \beta v_2 + \tfrac{\gamma}{k} \sin kv_3 + \tfrac{\delta}{k}(1-\cos kv_3)\\ -v_3\end{array}\right) & \mbox{ when } \epsilon=1.\end{array}\right.\end{eqnarray}

\subsection{The Lie groups $\mbox{Aut}(\mathfrak{s}_2)$ and $\mbox{Aut}(S_2)$}

The automorphisms of $\mathfrak{s}_2$ and $S_2$ as computed in previous sections form Lie groups $\mbox{Aut}(\mathfrak{s}_2)$ and $\mbox{Aut}(S_2)$ respectively under composition of mappings. In this section we show that each automorphism of $\mathfrak{s}_2$ or $S_2$ is a composition of automorphisms in various subgroups of $\mbox{Aut}(\mathfrak{s}_2)$ or $\mbox{Aut}(S_2)$. Moreover, the groups $\mbox{Aut}(\mathfrak{s}_2)$ and $\mbox{Aut}(S_2)$ are isomorphic.

The elements of $\mbox{Aut}(\mathfrak{s}_2)$ with respect to the basis $\{\bm f_1, \bm f_2, \bm f_3\}$ are
\begin{equation} \label{eq:auts2}
\begin{array}{l}\mbox{Aut}(\mathfrak{s}_2) = \left\{ L = \left( \begin{array}{ccc} 0 & 1 & 0 \\ 1 & 0 & 0 \\ 0 & 0 & -1 \end{array}\right)^\epsilon\!\!\left(\! \begin{array}{ccc} \alpha & \beta & \gamma \\ -\beta & \alpha & \delta \\ 0 & 0 & 1 \end{array}\right)\!\!, \; \epsilon \in \{0,1\}, \; \alpha, \beta, \gamma, \delta \in \mathbb{R},\; \alpha^2 +\beta^2 \neq 0.\right\}
\end{array}
\end{equation}
Define the following subgroups of $\mbox{Aut}(\mathfrak{s}_2)$:
\[P:=\left\{ L \in \mbox{Aut}(\mathfrak{s}_2) \ : \ L= \left( \begin{array}{ccc} 0 & 1 & 0 \\ 1 & 0 & 0 \\ 0 & 0 & -1 \end{array}\right)^\epsilon, \ \epsilon \in \{0,1\}\right\}\]
\[R:=\left\{ L \in \mbox{Aut}(\mathfrak{s}_2) \ : \ L= \left(\! \begin{array}{ccc} \alpha & \beta & \gamma \\ -\beta & \alpha & \delta \\ 0 & 0 & 1 \end{array}\right), \  \ \alpha, \beta, \gamma, \delta \in \mathbb{R},\ \ \alpha^2 +\beta^2 \neq 0\right\}.\]
Then, noting that $R$ is normal in $\mbox{Aut}(\mathfrak{s}_2)$ and $\mbox{Aut}(\mathfrak{s}_2)= P R$, if $L \in \mbox{Aut}(\mathfrak{s}_2)$ then $L$ can be written uniquely as a product of an element of $P$ and an element of $R$. Also define the following subgroups of $R$:
\[S:=\left\{ L \in \mbox{Aut}(\mathfrak{s}_2) \ : \ L= \left( \begin{array}{ccc} \alpha & \beta & 0 \\ -\beta & \alpha & 0 \\ 0 & 0 & 1 \end{array}\right), \  \ \alpha, \beta \in \mathbb{R},\ \ \alpha^2+\beta^2 \neq 0\right\},\]
\[T:=\left\{ L \in \mbox{Aut}(\mathfrak{s}_2) \ : \ L= \left( \begin{array}{ccc} 1 & 0 & \gamma \\ 0 & 1 & \delta \\ 0 & 0 & 1 \end{array}\right), \  \ \gamma, \delta \in \mathbb{R}\right\}.\]
Then $S$ is normal in $R$ and $R=TS$ so that any $L\in \mbox{Aut}(\mathfrak{s}_2)$ can be uniquely written as a product of an element of $P$, an element of $T$ and an element of $S$ or $\mbox{Aut}(\mathfrak{s}_2) = PTS$.

There is a one to one correspondence between the automorphisms of $\mathfrak{s}_2$ and the automorphisms of $S_2$, in fact there is an isomorphism $\bm \mu:\mbox{Aut}(\mathfrak{s}_2) \to \mbox{Aut}(S_2)$, and the Lie group automorphism $\bm \mu(L) = \bm \phi_L$ corresponding to the Lie algebra automorphism $L=pts$, $p\in P, t \in T, s\in S$, can be uniquely decomposed as
\[\bm \mu (L) = \bm \mu (p) \circ \bm \mu(t) \circ \bm \mu (s)\]
where $\bm \mu(p) \in \bm \mu(P)$ and $\bm \mu(P)$ is a subgroup of $\mbox{Aut}(S_2)$, and so on.

\section{Discrete subgroups of $S_2$ and their symmetries}\label{sec:discrete}

Recall that in this paper we are investigating whether or not automorphisms of discrete subgroups of the Lie group $S_2$ extend to automorphisms of the continuous Lie group. The automorphisms of a discrete subgroup $D$ correspond to a certain subset of the geometrical symmetries of $D$, where the geometrical symmetries of $D$ are the changes of generators that preserve the set of points in $D$. In this section we introduce the discrete subgroups $D$ of $S_2$ and recall from Nicks and Parry \cite{Nicks1}--\cite{Nicks3} how different sets of generators of $D$ are related to each other, thus determining the geometrical symmetries of $D$.

\subsection{The discrete subgroups of $S_2$}

According to Auslander, Green and Hahn \cite{Auslander}, when $\theta \in SL_2(\mathbb{Z})$ has $\mbox{tr}(\theta) \in \{ -2,-1,0,1\}$ the discrete subgroups $D \subset S_2$ are isomorphic (via $r_m(\cdot)$ defined in \eqref{eq:elements}) to a discrete subgroup $D_m$ of $S_m$ and $D_m$ is generated by three elements

\begin{equation}\label{eq:generators}
    A\equiv r_m (\bm e_3)= \left( \begin{array}{cc}
    \\[-0.3cm]
    \multicolumn{2}{c}{\theta}\\[0.3cm]
\hline
0&0\\
0&0\end{array}\begin{array}{|cc}
0&0\\0&0\\\hline1&1\\0&1\end{array}\right), B\equiv r_m (\bm e_1) = \left( \begin{array}{cccc}
1&0&0&1\\
0&1&0&0\\
0&0&1&0\\
0&0&0&1\end{array}\right), C\equiv r_m(\bm e_2) = \left( \begin{array}{cccc}
1&0&0&0\\
0&1&0&1\\
0&0&1&0\\
0&0&0&1\end{array}\right).
  \end{equation} Note that here we are again working with respect to the basis $\{\bm e_1, \bm e_2, \bm e_3\}$. Let $(X,Y)=X^{-1} Y^{-1} XY$ denote the commutator of elements $X$, $Y \in S_m$ where juxtaposition denotes matrix multiplication. We then note that
  \begin{equation}\label{eq:Dcomms}
    (A,B)=B^{1-d} C^c,\quad  (A,C)= B^bC^{1-a},\quad (B,C)=\mathbb{I}_4,
  \end{equation}
  where $\mathbb{I}_4$ is the identity element in $D_{m}$. From this one can see that any element of $D_{m}$ can be expressed as a product of the form
  \[
  d_m=A^{\alpha_1}B^{\beta_1}C^{\gamma_1}A^{\alpha_2}B^{\beta_2}C^{\gamma_2}\dots A^{\alpha_r}B^{\beta_r}C^{\gamma_r} = A^{\alpha_1+\cdots\alpha_r}B^MC^N
  \]
  for some $M,N \in \mathbb{Z}$ where $\alpha_i, \beta_i, \gamma_i \in \mathbb{Z}$, $i=1,2,\ldots,r$ and $r\in \mathbb{Z}$. A general element $d_m =A^QB^M C^N \in D_{m}$, $Q,M,N \in \mathbb{Z}$ has the representation
  \begin{equation}\label{eq:Delement}
    d_m= \left( \begin{array}{cc}
    \\[-0.3cm]
    \multicolumn{2}{c}{\theta^Q}\\[0.3cm]
\hline 0&0\\
0&0\
\end{array}\begin{array}{|ccc}
\begin{array}{c} 0\\0 \end{array} & \theta^Q \left(\begin{array}{c} M\\N \end{array}\right)\\\hline1&Q\\0&1\end{array}\right) = r_m(\bm x), \quad \mbox{where} \ \ \bm x= \left( \begin{array}{c} \theta^Q \left(\begin{array}{c} M\\N \end{array}\right) \\ Q\end{array}\right)\in S_2,\end{equation}
with respect to the basis $\{ \be_1, \be_2, \be_3\}$.
\par
It is then clear that since $\theta \in SL_2(\mathbb{Z})$, $r_m^{-1}(D_{m}) = D = (\mathbb{Z}^3, \bm \psi)$. Thus the discrete structures which are the analogues of perfect lattices $L$ in this case are the lattice $\mathbb{Z}^3$ with group multiplication $\bm \psi$ given by \eqref{eq:groupmult}.

\subsection{Symmetries of $D$}\label{sec:symms}

The symmetries of the discrete subgroup $D \subset S_2$ are the choices of three elements $\bg_1, \bg_2, \bg_3 \in D$  such that the subgroup of $D$ generated by the three elements (which we shall denote $G=gp\{\bg_1, \bg_2, \bg_3\}$) is in fact equal to $D$. These changes of generators preserve the integer lattice $\mathbb{Z}^3$. For the discrete subgroups of $S_2$, the conditions on $\bg_1, \bg_2, \bg_3$ that are necessary and sufficient that $G=D$ can be shown to be precisely the conditions that are necessary and sufficient for the commutator subgroups $G^\prime=(G,G)$ and $D^\prime=(D,D)$ to be equal (see Nicks and Parry \cite{Nicks2}). In this section we state these conditions without proof. Proofs of the statements below can be found in \cite{Nicks2} or \cite{Nicks3}.

Let $\bg_1, \bg_2, \bg_3$ be elements of $D$. That is, $\bg_{im}:=r_m(\bg_i)$ is a word in the generators $A, B, C$ of $D_m$ and we can use the commutator relations \eqref{eq:Dcomms} to write
\begin{equation}\label{eq:generatorsG}
    \begin{array}{rcl}
    \bm g_{1m}&=& A^{\alpha_1}B^{\beta_1}C^{\gamma_1},\\
     \bm g_{2m}&=& A^{\alpha_2}B^{\beta_2}C^{\gamma_2},\qquad \alpha _i, \beta_i, \gamma_i \in \mathbb{Z},\; i=1,2,3.\\
      \bm g_{3m}&=& A^{\alpha_3}B^{\beta_3}C^{\gamma_3},\end{array}
 \end{equation}

 If $G=gp\{\bg_1, \bg_2, \bg_3\}=D$ then clearly we must have \[G_m= gp\{\bg_{1m}, \bg_{2m}, \bg_{3m}\}=D_m.\] Since $A \in D_m$ it must also be the case that $A \in G_m$ if we are to have $G=D$. Due to the particular form of the commutator relations \eqref{eq:Dcomms} for the generators $A,B,C$ of $D_m$ this implies that $\hbox{hcf}(\alpha_1, \alpha_2, \alpha_3)=1$. In that case, the following lemma holds.

 \begin{lem}\label{lem:change}
Let $\bm g_{1m}, \bm g_{2m},\bm g_{3m}$ be given by \eqref{eq:generatorsG}, let $G_m=gp(\bm g_{1m}, \bm g_{2m},\bm g_{3m})$ and suppose that $\mathop{\rm hcf}(\alpha_1,\alpha_2,\alpha_3)=1$.  Then there is a set of generators of $G_m$, denoted  $\bm g_{1m}', \bm g_{2m}',\bm g_{3m}'$, such that
\begin{equation}\label{eq:equivgen}
\!\bm g_{1m}'=AB^{\beta_1'}C^{\gamma_1'}, \ \bm g_{2m}'=B^{\beta_2'}C^{\gamma_2'},\ \bm g_{3m}'=B^{\beta_3'}C^{\gamma_3'}, \ \beta_i', \gamma_i'\in \mathbb{Z}, \ i=1,2,3.\!\!\!
\end{equation}
\end{lem}

See Nicks and Parry \cite{Nicks3} for the proof of this lemma. It is then shown in Nicks and Parry \cite{Nicks1} that if one defines  $\bm \tau_1, \bm \tau_2, \bm \tau_3, \bm \tau_4 \in \mathbb{Z}^2$ by
\begin{equation}\label{5.10}
\bm \tau_1=\left(\begin{array}{c} \beta_2'\\\gamma_2'\end{array}\right),
\bm \tau_2=\left(\begin{array}{c} \beta_3'\\\gamma_3'\end{array}\right),
\bm \tau_3=\theta \left(\begin{array}{c} \beta_2'\\\gamma_2'\end{array}\right),
\bm \tau_4=\theta \left(\begin{array}{c} \beta_3'\\\gamma_3'\end{array}\right);
\end{equation}
where the values of $\beta_2', \beta_3', \gamma_2',\gamma_3'$ are as in \eqref{eq:equivgen}, then conditions necessary and sufficient that $G=D$ are that
\begin{eqnarray}\label{5.11}
 &&\mbox{hcf} \left(\tau_{11}, \tau_{12}, \tau_{13}, \tau_{14}\right)=\mbox{hcf}\left(\tau_{21}, \tau_{22}, \tau_{23}, \tau_{24}\right)=1\\ \label{5.12}&&\mbox{hcf} \left(\left\{ \bm \tau_i\wedge\bm \tau_j; i<j, i,j=1,2,3,4\right\}\right)=1,
\end{eqnarray} where the components of $\bm \tau_i,\; i=1,2,3,4$, are $\bigl(\begin{smallmatrix}\tau_{1i}\\ \tau_{2i}\end{smallmatrix} \bigr)$. From now on we will drop the primes on the $\beta_i'$, $\gamma_i'$ in the definitions of $\bm \tau_j$, $j=1,2,3,4$.

\section{Automorphisms of discrete subgroups} \label{sec:automorphisms}

In the previous section we gave details of the possible changes of generators of a discrete subgroup $D \subset S_2$. These are the symmetries of $D$. We now consider which of these symmetries extend to automorphisms of the discrete subgroup $D$.

\subsection{Changes of generators which extend to automorphisms of $D$}\label{sec:auto}

Here we state a result (Lemma \ref{lem:auto}) that gives necessary and sufficient conditions for a change of generators of $D$ to extend to an automorphism of $D$. The proof of the lemma is given in Nicks and Parry \cite{Nicks3}, and follows from results of Johnson \cite{Johnson} and Magnus, Karrass and Solitar \cite{Magnus} concerning free substitutions and automorphisms.

Let $\bg_{1m}, \bg_{2m}, \bg_{3m} \in D_m$ as in \eqref{eq:generatorsG} satisfy the conditions stated in the previous section so that they generate the group $D_m$ (which is isomorphic to $D$). Recall that $A,B,C$ given by \eqref{eq:generators} also generate $D_m$. Thus $A,B$ and $C$ can each be written as a word in $\bg_{1m}, \bg_{2m}, \bg_{3m}$ and their inverses. Moreover, the commutator relations \eqref{eq:Dcomms} can be expressed as relations in terms of the generators $\bg_{1m}, \bg_{2m}, \bg_{3m}$. We define mutually inverse mappings $\bm\phi, \bm\tau$ between the sets of generators $\{A,B,C\}$, $\{\bm g_{1m}, \bm g_{2m}, \bm g_{3m}\}$ by
\begin{equation}\label{eq:inverses}
   \begin{array}{rcl}
  \bm \phi(A) &=&  \bm g_{1m},\\
     \bm \phi(B) &=&  \bm g_{2m},\\
      \bm \phi(C) &=&  \bm g_{3m},\end{array}  \qquad \qquad  \begin{array}{rcl}
  \bm \tau(\bm g_{1m}) &=&  A,\\
     \bm \tau(\bm g_{2m}) &=&  B,\\
      \bm \tau(\bm g_{3m}) &=&  C,\end{array}
\end{equation} and the following lemma holds.

\begin{lem}\label{lem:auto}Let the mappings $\bm \phi$ and $\bm \tau$ be as defined in \eqref{eq:inverses}. These mappings extend to mutually inverse automorphisms $\bm \phi^\prime$, $\bm \tau^\prime$ of $D_m$ if
\begin{enumerate}
\item[(i)] the commutator relations \eqref{eq:Dcomms} continue to hold when $A$, $B$ and $C$ are replaced by $\bm\phi(A)$, $\bm \phi(B)$ and $\bm \phi(C)$ respectively; and
\item[(ii)] the relations in terms of the generators $\bg_{1m}, \bg_{2m}, \bg_{3m}$ obtained from the commutator relations \eqref{eq:Dcomms} continue to hold when $\bm g_{1m},  \bg_{2m}, \bg_{3m}$ are replaced by $\bm \tau( \bg_{1m}), \bm \tau( \bg_{2m}), \bm \tau( \bg_{3m})$ respectively.
\end{enumerate}
Conversely, if $\bm \phi^\prime$ and $\bm \tau^\prime= (\bm \phi^\prime)^{-1}$ are automorphisms of $D_m$ then conditions (i) and (ii) hold, and in addition the commutator relations \eqref{eq:Dcomms} continue to hold when $A$, $B$ and $C$ are replaced by $(\bm \phi^\prime)^{-1}(A)$, $(\bm \phi^\prime)^{-1}(B)$ and $(\bm \phi^\prime)^{-1}(C)$ respectively.
\end{lem}

See Nicks and Parry \cite{Nicks3} for the proof of this lemma. We now use this result to compute the automorphisms of $D_m$. Suppose that $\bm \phi$ and $\bm \tau$ are the changes of generators of $D_m$ defined in \eqref{eq:inverses} and they satisfy the conditions in section \ref{sec:symms}. Furthermore, let
\begin{equation}\label{eq:tau} \bm \tau(A) =A^{p_1}B^{q_1}C^{r_1}, \qquad  \bm \tau(B) =A^{p_2}B^{q_2}C^{r_2}, \qquad  \bm \tau(C) =A^{p_3}B^{q_3}C^{r_3}\end{equation} for  $p_i,q_i,r_i \in \mathbb{Z}$, $i=1,2,3$.

If $\bm \phi$ is to extend to an automorphism $\bm \phi^\prime$ of $D_m$ then by Lemma \ref{lem:auto} and the third commutator relation of \eqref{eq:Dcomms} the mapping must satisfy
\[(\bm\phi(B), \bm \phi(C)) = \mathbb{I}_4, \] which expresses the fact that $\bm\phi(B)$ and $\bm\phi(C)$ must commute. This implies that we must have $\alpha_2=\alpha_3=0$ in \eqref{eq:generatorsG}. Similarly we deduce that $p_2=p_3=0$ in \eqref{eq:tau} if $\bm \tau$ extends to an automorphism $\bm \tau^\prime$. Since $\bm \phi^\prime$ and $\bm \tau^\prime$ are to be mutually inverse automorphisms $\bm \tau^\prime\circ \bm \phi^\prime$ must be the identity and therefore $\bm \tau(\bm \phi(A))=A$, $\bm \tau(\bm \phi(B))=B$, $\bm \tau(\bm \phi(C))=C$, from which we deduce that
\begin{equation}\label{eq:autoconds}\alpha_1=p_1=\zeta=\pm 1 \qquad \mbox{ and } \qquad \left( \begin{array}{cc} q_2 &q_3\\ r_2 & r_3 \end{array}\right)\left( \begin{array}{cc} \beta_2  &\beta_3\\ \gamma_2 & \gamma_3 \end{array}\right)= \left( \begin{array}{cc} 1 &0\\ 0 & 1 \end{array}\right).\end{equation} Thus
\begin{equation}\label{eq:chi}
\chi := \left( \begin{array}{cc} \beta_2  &\beta_3\\ \gamma_2 & \gamma_3 \end{array}\right)\in GL_2(\mathbb{Z}),
\end{equation} and $\bigl(\begin{smallmatrix}q_2&q_3\\ r_2&r_3\end{smallmatrix} \bigr)= \chi^{-1}$.
Finally, from the first two commutator relations of \eqref{eq:Dcomms} we must have
\[
(\bm\phi(A), \bm \phi(B)) = \bm \phi(B)^{1-d} \bm \phi(C)^c \qquad \mbox{and} \qquad (\bm \phi(A), \bm \phi(C)) =\bm  \phi(B)^{b} \bm \phi(C)^{1-a},
\]
which can be shown to imply that the matrix $\chi$ defined in \eqref{eq:chi} must satisfy
\begin{equation}\label{eq:thetasym}
    \theta^{\zeta} \chi = \chi \theta \quad\text{ for }\quad\zeta=\alpha_1=\pm 1,
\end{equation} for the given matrix $\theta\in SL_2(\mathbb{Z})$. Further details of these computations are given in Nicks and Parry \cite{Nicks3}.

Recall that the discrete subgroup $D_m$ depends on a given matrix $\theta \in SL_2(\mathbb{Z})$. The conditions that a change of generators $\bm \phi$ of $D_m$ extends to an automorphism of $D_m$ can be summarized as
\begin{equation}\label{eq:summary} \bm \phi(A) = A^\zeta B^{\beta_1} C^{\gamma_1}, \qquad  \bm \phi(B) = B^{\beta_2} C^{\gamma_2},\qquad   \bm \phi(C) = B^{\beta_3} C^{\gamma_3},\end{equation}
where $\zeta=\pm1$, $\beta_1, \gamma_1$ are arbitrary integers and the matrix $\chi$ of the exponents $\beta_i, \gamma_i$, $i=2,3$, defined in \eqref{eq:chi}, satisfies condition \eqref{eq:thetasym}. These conditions are also sufficient that the conditions of Lemma \ref{lem:auto} hold. Hence it remains to determine the matrices $\chi$ satisfying \eqref{eq:thetasym}.

\subsection{Computing the automorphisms of $D$}

By the results obtained in section \ref{sec:auto}, in order to determine the changes of generators $\bm \phi$ which extend to automorphisms of $D_m$ it remains to compute the $2 \times 2$ matrices of exponents, denoted $\chi$ above, that satisfy condition \eqref{eq:thetasym} for a given matrix $\theta \in SL_2(\mathbb{Z})$. We now summarize here the work of Baake and Roberts \cite{Baake} on how to determine the matrices $\chi \in GL_2(\mathbb{Z})$ satisfying \eqref{eq:thetasym} for a given matrix $\theta \in SL_2(\mathbb{Z})$ with $\mbox{tr}(\theta) \in \{-2,-1,0,1\}$.

As in \cite{Baake}, we define the set of \textit{symmetries} of the matrix $\theta \in SL_2(\mathbb{Z})$ as
 \begin{equation} \label{6.16}
 S(\theta):= \{ \chi \in GL_2(\mathbb{Z}) \ : \ \theta\chi =\chi\theta\}.
 \end{equation}
 This is a subgroup of $GL_2(\mathbb{Z})$ and is the centralizer of $\theta$ in $GL_2(\mathbb{Z})$. If $\chi \in GL_2(\mathbb{Z})$ satisfies $\chi \theta \chi^{-1}= \theta^{-1}$ then we say that $\chi$ is a reversing symmetry of $\theta$ and when such a $\chi$ exists we call $\theta$ reversible. We define the following subgroup of $GL_2(\mathbb{Z})$ as the reversing symmetry group of $\theta$,
  \begin{equation} \label{6.17}
  R(\theta):= \{ \chi \in GL_2(\mathbb{Z}) \ : \ \chi \theta \chi^{-1}= \theta^{\pm 1}\}.
  \end{equation}
  It is a subgroup of the normalizer of the group generated by $\theta$ in $GL_2(\mathbb{Z})$ and clearly $S(\theta)\subset R(\theta)$. (If $H\subset G$, the normalizer of $H$ in $G$ is $\left\{ a\in G: aH=Ha\right\}$). Moreover, $S(\theta)$ is a normal subgroup of $R(\theta)$.

Given a matrix $\theta \in SL_2(\mathbb{Z})$ with $\mbox{tr}(\theta) \in \{-2,-1,0,1\}$ we want to compute $R(\theta)$. As observed earlier, when $\mbox{tr}(\theta)=-2$, $\theta$ can only lie on a one-parameter subgroup of $SL_2(\mathbb{R})$ if $\theta=-\mathbb{I}_2$. Clearly, in this case $R(\theta)$ is $GL_2(\mathbb{Z})$. If $\mbox{tr}(\theta) \in \{-1,0,1\}$ then there are potentially two possibilities
\begin{enumerate}
   \item[(i)] $R(\theta) = S(\theta)$. This occurs in cases where $\theta$ is not reversible;
   \item[(ii)] The index of $S(\theta)$ in $R(\theta)$ is $2$ so that $R(\theta)$ is a $C_2$-extension of $S(\theta)$ (where $C_2$ is the cyclic group of order $2$).
   \end{enumerate}
For a given matrix $\theta$ with $\mbox{tr}(\theta) \in \{-1,0,1\}$ there is a finite algorithm for computing $R(\theta)$ which we briefly summarize following Baake and Roberts \cite{Baake}.

 \subsubsection{Computing $S(\theta)$}\label{sec:Stheta}

 Clearly $\pm\theta^m\in S(\theta)$ for all $m\in \mathbb{Z}$ and we can observe that
  \begin{itemize}
 \item if $\mbox{tr}(\theta)=-1$ then $\theta^3=\mathbb{I}_2$;
 \item if $\mbox{tr}(\theta)=0$ then $\theta^4=\mathbb{I}_2$;
 \item and if $\mbox{tr}(\theta)=1$ then $\theta^6=\mathbb{I}_2$.
 \end{itemize} Does $S(\theta)$ contain any other matrices $\chi$?

When $\mbox{tr}(\theta) \in \{-1,0,1\}$ the eigenvalues of $\theta$, $\lambda$ and $1/\lambda$, are distinct. Consequently $\theta$ can be diagonalised by a matrix $U$ which has entries in $\mathbb{Q}(\lambda)$, the smallest field extension of the rationals that contains $\lambda$, so that
\[
U^{-1} \theta U= \left(\begin{array}{cc}\lambda & 0 \\ 0 & \frac{1}{\lambda}\end{array}\right) \qquad \mbox{where} \ \theta = \left(\begin{array}{cc}a & b \\ c & d\end{array}\right), \ \ \ U= \left(\begin{array}{cc} b & b   \\ \lambda -a & \frac{1}{\lambda}-a\end{array}\right).
\]
Suppose that $\chi \in S(\theta)$. Then $U^{-1} \chi U$ commutes with $U^{-1} \theta U$ and since only diagonal matrices can commute with a diagonal matrix with different diagonal entries,
\[
U^{-1} \chi U= \left(\begin{array}{cc}\mu_1 & 0 \\ 0 & \mu_2\end{array}\right), \qquad \mbox{some $\mu_1$, $\mu_2 \in \mathbb{Q}(\lambda)$}.
\]
Thus $S(\theta)$ contains all matrices $\chi \in GL_2(\mathbb{Z})$ which are diagonalised by $U$. Since $\chi \in GL_2(\mathbb{Z})$, it has eigenvalues $\mu_1, \mu_2$ which are algebraic integers. Also $\mu_2 = \pm \mu_1^{-1}$ and hence $\mu_1$ and $\mu_2$ are units in $\mathcal{O}$, the maximal order of $\mathbb{Q}(\lambda)$ (i.e the intersection of $\mathbb{Q}(\lambda)$ with the set of algebraic integers). Thus $S(\theta)$ is isomorphic to a subgroup of the unit group of $\mathcal{O}$.
\begin{itemize}
\item When $\mbox{tr}(\theta)=0$, $\mathbb{Q}(\lambda) = \mathbb{Q}(\sqrt{-1})$ and the unit group is isomorphic to $\{\pm \mathbb{I}_2, \pm \theta\}$. Since $\theta$ commutes with each of these matrices, $S(\theta)=\{\pm \mathbb{I}_2, \pm \theta\}\simeq C_4$.
\item When $\mbox{tr}(\theta)=\pm 1$, $\mathbb{Q}(\lambda) = \mathbb{Q}(\sqrt{-3})$ and the unit group is isomorphic to $\{\pm \mathbb{I}_2, \pm \theta, \pm \theta^2\}$. Since $\theta$ commutes with each of these matrices, $S(\theta)=\{\pm \mathbb{I}_2, \pm \theta, \pm \theta^2\}\simeq C_6$ (see \cite{Borevich}, \cite{Cohn}).
\end{itemize}

In particular, when $\mbox{tr}(\theta) \in \{-1,0,1\}$ the group of matrices $S(\theta)$ which commute with $\theta$ is finite.

 \subsubsection{Computing $R(\theta)$}\label{sec:Rtheta}
 In order to compute $R(\theta)$ for a given matrix $\theta$ we need to determine the matrices $\Lambda\in GL_2(\mathbb{Z})$ such
that $\Lambda\theta\Lambda^{-1} =\theta^{-1}$. Recall that either $R(\theta) = S(\theta)$ (if no such
$\Lambda$ exists) or $R(\theta)$ is a $C_2$-extension of $S(\theta)$. Also, if $\theta$ has a reversing symmetry $\Lambda$ then all other
reversing symmetries of $\theta$ are obtained as $\Lambda\chi$ where $\chi \in S(\theta)$.

For any matrix $\theta \in SL_2(\mathbb{Z})$, $\mbox{tr}(\theta)=\mbox{tr}(\theta^{-1})$. Moreover it can be shown that when $\mbox{tr}(\theta) \in \{-1,0,1\}$, $\theta$ is conjugate in $GL_2(\mathbb{Z})$ to
\[ \left(\begin{array}{cc}\mbox{tr}(\theta) & 1 \\ -1 & 0\end{array}\right).\] That is, there is only one conjugacy class of matrices for each value of $\mbox{tr}(\theta)\in \{-1,0,1\}$ and hence $\theta \in SL_2(\mathbb{Z})$ must be conjugate in $GL_2(\mathbb{Z})$ to its inverse. In other words, for any $\theta \in SL_2(\mathbb{Z})$ with $\mbox{tr}(\theta)\in \{-1,0,1\}$ there exists a matrix $\Lambda \in GL_2(\mathbb{Z})$ such that $\Lambda\theta\Lambda^{-1} =\theta^{-1}$. Hence $\theta$ has a reversing symmetry $\Lambda$ and $R(\theta)$ is a $C_2$-extension of $S(\theta)$. This implies that
\[ R(\theta) = \left\{ \begin{array}{ll} D_4= \{ \pm \mathbb{I}_2, \pm \theta, \pm \Lambda, \pm \Lambda\theta\} & \mbox{if $\mbox{tr}(\theta)=0$}\\ D_6 =\{ \pm \mathbb{I}_2, \pm \theta, \pm \theta^2  \pm \Lambda, \pm \Lambda\theta, \pm \Lambda \theta^2\} & \mbox{if $\mbox{tr}(\theta)=\pm 1$.}\end{array} \right.\]

Notice that for matrices $\theta \in SL_2(\mathbb{Z})$ with $\mbox{tr}(\theta)\in \{-1,0,1\}$ there is only a finite number of symmetries and reversing symmetries of $\theta$ but the number of changes of generators $\bm \phi$ which extend to automorphisms of the corresponding discrete subgroup $D_m$ is infinite due to the fact that the choice of the exponents $\beta_1$ and $\gamma_1\in \mathbb{Z}$ in \eqref{eq:summary} is arbitrary.

Also, note that if $\chi \theta= \theta^\zeta \chi$ for a given $\theta \in SL_2(\mathbb{Z})$ with $\mbox{tr}(\theta) \in \{-1,0,1\}$ then the entries $\beta_3$, $\gamma_3$ in the matrix $\chi$ can be expressed in terms of the entries $\beta_2$, $\gamma_2\in \mathbb{Z}$ and the entries $a,b,c,d \in \mathbb{Z}$ of $\theta$.

\section{Extensions of automorphisms} \label{sec:extensions}

We next discuss which of the automorphisms of $D$ computed in section \ref{sec:automorphisms} extend to automorphisms of the continuous Lie group $S_2$. Recall that this enables us to classify the symmetries (changes of generators) $\bm \phi$ of $D$ given in section \ref{sec:symms} as
\begin{enumerate}
\item `Elastic' if $\bm \phi$ extends to an automorphism $\bm \phi^\prime$ of $D$ and $\bm \phi^\prime$ extends uniquely to an automorphism $\widetilde{\bm \phi}:S_2 \to S_2$. These changes of generators are restrictions of elastic deformations of the continuum crystal whose distribution of defects is uniform and has corresponding Lie group $S_2$.
\item `Inelastic' if \begin{enumerate}
                    \item if $\bm \phi$ does not extend to an automorphism of $D$; or
                    \item if $\bm \phi$ extends to an automorphism $\bm \phi^\prime$ of $D$ but $\bm \phi^\prime$ does not extend uniquely to an automorphism of $S_2$.
                    \end{enumerate} These changes of generators preserve the set of points in $D$ but not its group structure.
\end{enumerate}

In this section we will show by direct calculation that for crystals with underlying Lie group $S_2$ there are no changes of generators $\bm \phi$ in the class 2(a) above. That is; all automorphisms of $D\subset S_2$ extend uniquely to automorphisms of $S_2$. This is a computation which has not been required in the analysis of other classes of crystals with uniform distributions of defects, where the underlying Lie group if nilpotent or in the solvable class $S_1$. In those cases results of Mal'cev \cite{Malcev} and Gorbatsevich \cite{Gorbatsevich} respectively guarantee that any automorphism of a uniform discrete subgroup   will extend uniquely to an automorphism of the ambient continuous Lie group. Since no corresponding general result exists for automorphisms of the class of solvable Lie group $S_2$ we proceed by direct calculation. We show explicitly how each of the automorphisms of $D$ computed in section \ref{sec:automorphisms} extends uniquely to an automorphisms of $S_2$ computed in section \ref{sec:S2auto}.

 We should note first that a particular discrete subgroup $D$ (determined by the matrix $\theta \in SL_2(\mathbb{Z})$ with $\mbox{tr}(\theta) \in \{-2,-1,0,1\}$) is a subgroup of infinitely many (isomorphic) Lie groups $S_2$ corresponding to different choices of $k$ where $2\cos k = \mbox{tr}(\theta)$. (Recall that $k = \pm k_0 +2\pi n$ for any $n \in \mathbb{Z}$ where $k_0$ is the value of $k$ in \eqref{eq:k} with $n=1$, so there are infinitely many possible values of $k$ for a given value of $\mbox{tr}(\theta) \in \{-2,-1,0,1\}$. Choosing one of these values of $k$ determines the matrix $\mathcal{A}$ which specifies a particular Lie group in the isomorphism class of $S_2$.)

 Assume that we are given $\theta$ with $\mbox{tr}(\theta) \in \{-2,-1,0,1\}$ and a corresponding fixed value of $k$. Thus a particular group $S_2(k)$ and discrete subgroup $D=D(\theta)\subset S_2(k)$ are determined. Here we will show that the automorphisms of $D(\theta)$ extend uniquely to automorphisms of $S_2(k)$. Notice that in section \ref{sec:S2auto} we computed the automorphisms of $S_m$ with respect to the basis $\{\bm f_1, \bm f_2, \bm f_3\}$ and the automorphisms of $D_m$ in section \ref{sec:automorphisms} are given in terms of changes of generators expressed with respect to the basis $\{\bm e_1, \bm e_2, \bm e_3\}$. The change of basis is specified by the matrix $M$ in \eqref{eq:M} which depends on $k$.

 Let $\bm \phi_D:D \to D$ be an automorphism and suppose that it extends to an automorphism $\widetilde{\bm \phi}:S_2 \to S_2$. We show that this extension exists and is unique. Let $\bm x = x_i \bm e_i \in D$ and $r_m(\bm x)= A^qB^mC^n$ for some $q,m,n \in \mathbb{Z}$. Thus
 \[ \bm x =\left\{ \begin{array}{ll}  \left( \begin{array}{c} \theta^q\left(\begin{array}{c} m\\n\end{array}\right) \\ q\end{array}\right) & \mbox{with respect to the basis $\{\bm e_1, \bm e_2, \bm e_3\}$} \\  M^{-T} \left( \begin{array}{c} \theta^q\left(\begin{array}{c} m\\n\end{array}\right) \\ q\end{array}\right) & \mbox{with respect to the basis $\{\bm f_1, \bm f_2, \bm f_3\}$}\end{array} \right.\]
If $\bm \phi_D$ extends to $\widetilde{\bm \phi}$ then $\bm \phi_D(\bm x) = \widetilde{\bm \phi}(\bm x)$ for all $\bm x \in D$. In section \ref{sec:automorphisms} we computed all automorphisms $\bm \phi_{m}$ of $D_m$ and these automorphisms are in one to one correspondence with the automorphisms of $D$ since they satisfy
\[ \bm \phi_{m}(r_m(\bm x))= r_m(\bm \phi_D(\bm x)) \qquad \mbox{for $\bm x \in D$.}\]
Therefore
\begin{eqnarray} \bm \phi_D(\bm x) &=& r_m^{-1}(\bm \phi_m(r_m(\bm x))) = r_m^{-1}(\bm \phi_m (A^qB^mC^n))\nonumber \\
&=& r_m^{-1}( (A^\zeta B^{\beta_1} C^{\gamma_1})^q (B^{\beta_2} C^{\gamma_2})^m(B^{\beta_3} C^{\gamma_3})^n)\nonumber\\
&=& r_m^{-1}(A^{q\zeta} B^{s+m\beta_2 + n\beta_3} C^{t+m\gamma_2 + n\gamma_3})\nonumber\\
&=& \label{eq:autoD} M^{-T} \left( \begin{array}{c} \theta^{q\zeta}\left[\left(\begin{array}{c} s\\t\end{array}\right)+\chi \left(\begin{array}{c} m\\n\end{array}\right) \right] \\ q\zeta\end{array}\right),\end{eqnarray} with respect to the basis $\{\bm f_1, \bm f_2, \bm f_3\}$, recalling the definition of the matrix of exponents $\chi$ from \eqref{eq:chi} and where
\[ \left(\begin{array}{c} s\\t\end{array}\right) =\left\{ \begin{array}{ll} \left( \begin{array}{c} 0 \\ 0\end{array}\right) & \mbox{if $q=0$.}\\ \left(\sum_{j=0}^{q-1} \theta^{-j\zeta}\right)\left( \begin{array}{c} \beta_1 \\ \gamma_1\end{array}\right) & \mbox{if $q\neq 0$.}\end{array}\right.\]

From \eqref{eq:S2auto} the automorphism $\widetilde{\bm \phi}$ is uniquely determined by the values of $\epsilon \in \{0,1\}$ and $\alpha, \beta, \gamma, \delta \in \mathbb{R}$ and satisfies
\begin{equation}\label{eq:autoS} \widetilde{\bm \phi}(\bm x) = \left(\begin{array}{cc} W(\epsilon) \left(\begin{array}{cc} \alpha & \beta \\ -\beta &\alpha\end{array} \right) \overline{M}^{-T} \theta^q \left(\begin{array}{c} m \\ n\end{array}\right) + F(\mathcal{B}(-1)^\epsilon q) W(\epsilon) q\left(\begin{array}{c} \gamma \\ \delta \end{array}\right)  \\ (-1)^\epsilon q \end{array} \right)\end{equation}
where
\[ \overline{M} =\left( \begin{array}{cc} -b'(0) & a'(0)+k \\ -b'(0) & a'(0)-k \end{array} \right),\] (the top left $2 \times 2$ matrix contained in $M$) and the quantities $W(\epsilon)$ and $F(\mathcal{B}(-1)^\epsilon q)$ are as defined in section \ref{sec:S2auto}. Here the components are again given with respect to the basis $\{\bm f_1, \bm f_2, \bm f_3\}$.

Comparing \eqref{eq:autoD} and \eqref{eq:autoS} we see that if $\bm \phi_D$ is to extend to $\widetilde{\bm \phi}$, it must be the case that $\zeta = (-1)^\epsilon$ and furthermore, for all $q,m,n \in \mathbb{Z}$
\begin{eqnarray} \nonumber
&& \overline{M}^{-T} \theta^{q(-1)^\epsilon} \left[\left(\begin{array}{c} s\\t\end{array}\right)+\chi\left(\begin{array}{c} m\\n\end{array}\right) \right]= \\&& \qquad \qquad   W(\epsilon) \left(\begin{array}{cc} \alpha & \beta \\ -\beta &\alpha\end{array} \right) \overline{M}^{-T} \theta^q \left(\begin{array}{c} m \\ n\end{array}\right) + F(\mathcal{B}(-1)^\epsilon q) W(\epsilon) q\left(\begin{array}{c} \gamma \\ \delta \end{array}\right). \label{eq:extcond} \end{eqnarray}

Thus we must have
\begin{eqnarray}\label{eq:1} \overline{M}^{-T} \theta^{q(-1)^\epsilon} \chi &=& W(\epsilon)  \left(\begin{array}{cc} \alpha & \beta \\ -\beta &\alpha\end{array} \right) \overline{M}^{-T} \theta^q,\\ \label{eq:2}
\overline{M}^{-T} \sum_{j=1}^{q} \theta^{j(-1)^\epsilon}\left( \begin{array}{c} \beta_1 \\ \gamma_1\end{array}\right) &=& F(\mathcal{B}(-1)^\epsilon q) W(\epsilon) q\left(\begin{array}{c} \gamma \\ \delta \end{array}\right),\end{eqnarray} where $q\neq 0$ in \eqref{eq:2}.

Using the fact that $\theta^{(-1)^\epsilon}\chi=\chi\theta$ since $\bm \phi_D$ is an automorphism, from \eqref{eq:1} we find
\begin{equation} \left(\begin{array}{cc} \alpha & \beta \\ -\beta &\alpha\end{array} \right) = W(\epsilon)\overline{M}^{-T}\chi \overline{M}^{T} \end{equation} and hence the real numbers $\alpha$ and $\beta$ are uniquely determined by the values of $\epsilon$ and $\chi$.

Furthermore, let us define
\[ p:= \left\{ \begin{array}{ll} 2 & \mbox{when } \mbox{tr}(\theta)=-2\\ 3 & \mbox{when } \mbox{tr}(\theta)=-1\\  4 & \mbox{when } \mbox{tr}(\theta)=0\\ 6 & \mbox{when } \mbox{tr}(\theta)=1\\ \end{array}\right.\] Then we can note from \eqref{eq:F} that if $q \neq 0 \mod p$, the matrix $F(\mathcal{B}(-1)^\epsilon q)$ has a well defined inverse and we can write
\[ \left(\begin{array}{c} \gamma \\ \delta \end{array}\right) = \underbrace{\frac{1}{q} W(\epsilon) (F(\mathcal{B}(-1)^\epsilon q))^{-1}\overline{M}^{-T} \sum_{j=1}^{q} \theta^{j(-1)^\epsilon}}_{R(\epsilon)}\left( \begin{array}{c} \beta_1 \\ \gamma_1\end{array}\right)\] (If $q=0\mod p$ then both sides of \eqref{eq:2} are zero and we gain no information about $\gamma$ and $\delta$ for given values of $\beta_1$ and $\gamma_1$.) It can be observed that the value of the matrix $R(\epsilon)$ is independent of $q \neq 0 \mod p$ and is given by
\[R(\epsilon) = W(\epsilon) (F(\mathcal{B}(-1)^\epsilon))^{-T}\overline{M}^{-T}.\] Since this matrix is nonsingular, the values of $\gamma$ and $\delta$ are determined uniquely by the values of the exponents $\beta_1$ and $\gamma_1$ for a given automorphism of $D$. Thus we have shown directly that every automorphisms $\bm \phi_D:D \to D$ extends uniquely to an automorphism $\widetilde{\bm \phi}:S_2 \to S_2$ for a given value of $k$. In particular, we have shown that for a given automorphism $\bm \phi_D$ of $D$ specified by the matrix $\chi$, $\beta_1$, $\gamma_1 \in\mathbb{Z}$ and $\epsilon \in \{0,1\}$ there are unique values of $\alpha, \beta, \gamma, \delta$ and $\zeta$ which specify the automorphism $\widetilde{\bm \phi}:S_2 \to S_2$ to which $\bm \phi_D$ extends.

\section{Conclusions and discussion} \label{sec:conclusions}

The technical calculations given in this paper have allowed us to complete the classification of symmetries of crystals with uniform distributions of defects, which preserve the Lie group structure. Let us now briefly summarize the results.

The energy density $w$ of a crystal with a uniform distribution of defects depends on arguments $\{\bell_a\}= \{\bell_a(\bm 0)\}$ and $S$, where the vectors $\{\bell_a\}$ specify a set of generators of a discrete subgroup $D$ of the Lie group associated with the given constant value of the dislocation density tensor $S$. The underlying Lie group must be in one of three classes: it is either nilpotent or isomorphic to one of two classes of solvable Lie groups denoted $S_1$ and $S_2$. In this paper we have focused on the case where the Lie group is a solvable group in the class $S_2$.

A symmetry of the energy density function is a change of generators of $D$ which preserves the set of points in $\mathbb{R}^3$ associated with the elements of $D$. These symmetries have been shown to satisfy conditions \eqref{5.11}--\eqref{5.12} for subgroups $D \subset S_2$. Of these changes of generators, only those which extend to automorphisms $\bm \phi_D$ of $D$ also preserve the group structure. These are the changes of generators which satisfy condition \eqref{eq:thetasym}. These symmetries also have been shown in this paper to extend uniquely to automorphisms of the underlying continuous Lie group $S_2$ and hence they are (restrictions of) elastic deformations of the continuum defective crystal. We call such symmetries of $D$ `elastic'. The remaining symmetries of $D$ which do not extend to automorphisms of $D$ (or $S_2$) are classified as inelastic.

Together with work presented in \cite{Nicks3} and \cite{Parry10}, this completes the classification of all group preserving symmetries of uniform discrete defective crystals as either elastic or inelastic.

A future task will be to understand the properties of the inelastic symmetries with reference to corresponding mechanical problems including exploring any possible correlations between these symmetries and the presence of slip planes in crystalline materials.

\section*{Acknowledgements}
The author acknowledges the support of the UK Engineering and Physical Sciences Research Council through grant EP/G047162/1. The author wishes to thank Gareth Parry for many useful discussions and his advice concerning this paper.

\end{document}